\documentstyle[12pt]{article}

\newtheorem{remark}{Remark}

\begin{document}

\title{Decoherence and continuous measurements: \\
     Phenomenology and models}
\author{Michael B.~Mensky\\[1mm]
\small{P.N.Lebedev Physical Institute, 117924 Moscow, Russia}\\
\and 
\small{Fakult\"at f\"ur Physik der Universit\"at Konstanz}\\
\small{Postfach 5560 M 674, D-78434 Konstanz, Germany}}
\date{}

\maketitle              

\begin{abstract}
Decoherence of a quantum system induced by the interaction with its environment (measuring medium) may be presented phenomenologically as a continuous (or repeated) fuzzy quantum measurement. The dynamics of the system subject to continuous decoherence (measurement) may be described by the complex-Hamiltonian Schr\"odinger equation, stochastic Schr\"odinger equation of a certain type or (nonselectively) by the Lindblad master equation. The formulation of this dynamics with the help of restricted path integrals shows that the dynamics of the measured system depends only on the information recorded in the environment. With the help of the  complex-Hamiltonian Schr\"odinger equation, monitoring a quantum transition is shown to be possible, at the price of decreasing the transition probability (weak Zeno effect). The monitoring of the level transition may be realized by a long series of short weak observations of the system which resulting in controllable slow decoherence. 
\end{abstract}

\section{Models and phenomenology of decoherence}

Decoherence of a quantum system (appearance of classical features in its behavior) is caused by its interaction with the environment \cite{Zeh-bk96}. Decoherence may be presented as a series of fuzzy measurements or as a continuous fuzzy measurement (CFM). Hence, any phenomenological description of CFM may be considered as a phenomenological description of decoherence \cite{Men97rev,Men98revEng}. 

\subsection{Models for continuous fuzzy measurements (decoherence)}

Many models of decoherence in accordance with the scheme 

\quad

{\framebox{{\bf System}} $\leftrightarrow$ \framebox{Environment}}

\quad

\noindent
(i.e. including the system, its environment  and their interaction) were proposed  
\cite{Zeh70}-\cite{KonMenNamiot93}
(see also \cite{Zeh-bk96} and references therein). 
In the model of quantum diffusion or quantum Brownian motion \cite{CaldeiraLegg83} a particle moves through a crystal interacting with its atoms so that the information about its motion is recorded in the state of the crystal modes (phonons). The resulting equation for the density matrix of the particle has (in a special case) the form 
\begin{equation}
\dot\rho = -\frac{i}{\hbar} [H,\rho]
          - \frac 12 \kappa [\mathbf r\, [\mathbf r\, \rho]]
\label{quDiffusEq}\end{equation}
with $\kappa = 2\eta kT/\hbar^2$ ($\eta$ being the damping coefficient). Another model of quantum diffusion is constructed in \cite{KonMenNamiot93} as a model of continuous monitoring of the particle position due to its interaction with internal degrees of freedom of surrounding atoms. The resulting equation is the same but with the coefficient $\kappa = 2/\lambda^2\tau$ depending on the radius $\lambda$ of the particle-atom interaction and relaxation time $\tau$ of an atom. 

Both these models describe decoherence, but in the first case it is decoherence by crystal modes while in the second case --- decoherence by internal structure of atoms. The lesson which may be learnt is that decoherence may be treated as a continuous measurement (compare Eq.~(\ref{quDiffusEq}) with Eq.~(\ref{LindbladEq})). 

\subsection{Decoherence = continuous fuzzy measurement (CFM)}
\label{SectDecoCFM}

This feature is in fact general \cite{AudMen97En}: the process of gradual decoherence may be presented (or interpreted) as a process of quantum {\em continuous fuzzy measurement} (CFM). Indeed, an instantaneous (in reality short) fuzzy measurement is a measurement (entanglement) having a poor resolution, but in a series of such measurements the resolution improves, and finally becomes complete (in the case of a discrete spectrum of the measured observable). For example, a series of fuzzy measurements of an observable having two eigenvalues changes the state of the measured system in the following way: 
\begin{figure}[ht]
\let\picnaturalsize=N
\def\picsize{5cm}
\def\picfilename{rep-fuz.eps}
\ifx\nopictures Y\else{\ifx\epsfloaded Y\else\input epsf \fi
\let\epsfloaded=Y
\centerline{\ifx\picnaturalsize N\epsfxsize \picsize\fi \epsfbox{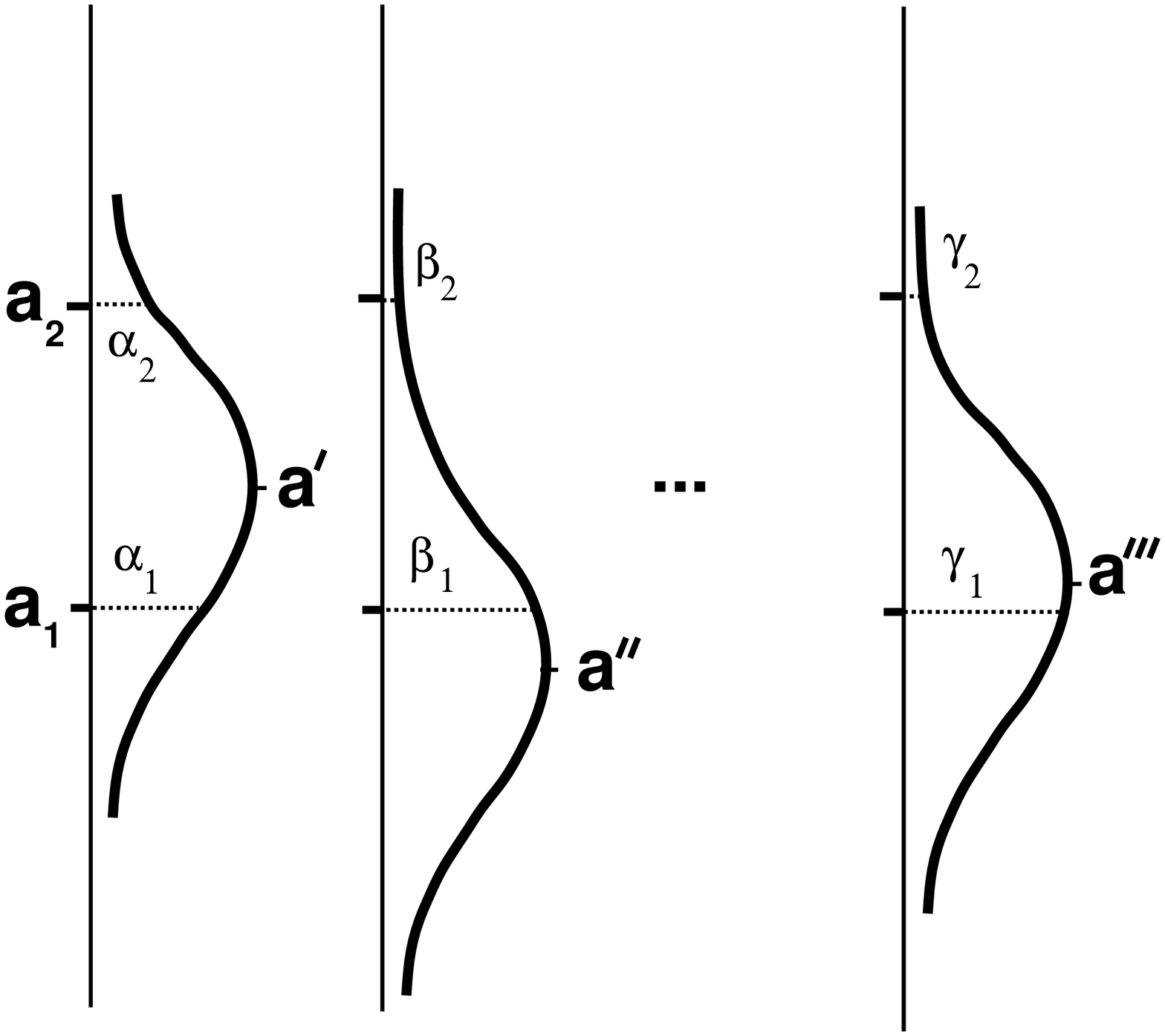}}}\fi
\label{fig-repeat-fuzzy}
\end{figure}
\begin{eqnarray*}
c_1|1\rangle+c_2|2\rangle &\rightarrow& \alpha_1 c_1|1\rangle+\alpha_2 c_2|2\rangle \\
	&\rightarrow& \beta_1\alpha_1 c_1|1\rangle+\beta_2\alpha_2 c_2|2\rangle \\
	& &\dots\dots\dots\dots\dots\dots\dots\\
	&\rightarrow& 
\gamma_1\dots\beta_1\alpha_1 c_1|1\rangle+\gamma_2\dots\beta_2\alpha_2 c_2|2\rangle\\
	&=& \tilde{c}_1|1\rangle \quad{\rm or}\quad \tilde{c}_2|2\rangle.
\end{eqnarray*}
The final result of this process may be presented by the von Neumann's projection, while the process itself is nothing else than gradual decoherence or CFM. 

\subsection{Phenomenology of CFM (open systems)} \label{SectPenomen}

One may describe CFM (decoherence) phenomenologically. This means that the environment is not explicitly included in the description, but its influence is taken into account phenomenologically, according to the scheme 

\quad

\framebox{{\bf System}} $\leftrightarrow$

\quad

\noindent
In this case the system is considered to be open. 

A {\em nonselective} description of an (open) continuously measured system is given by {\em Lindblad master equation} \cite{Lindblad76} which in the simplest case reads as follows (compare with Eq.~(\ref{quDiffusEq})):
\begin{equation}
\dot\rho = -\frac{i}{\hbar} [H,\rho]
          - \frac 12 \kappa [A, [A, \rho]]. 
\label{LindbladEq}\end{equation}
Here $A$ is an observable which is measured continuously and the constant $\kappa$ characterizes the measurement resolution (this interpretation was not given by Lindblad, but it is clear from another approach, see below Eq.~(\ref{complHamEq}) and Sect.~\ref{SectRPI}. ``Nonselective" means referring to no concrete state of the environment (equivalently, to no concrete measurement readout) but accounting for all of them with the corresponding probabilities. 

A {\em selective} description of the same process refers to a certain state of the environment or a certain measurement readout. In the simplest case, when the process is considered to be sequential (as in Sect.~\ref{SectDecoCFM}) rather than continuous, the evolution of the system is presented by the evolution operator 
$$
U_{\alpha}=
U(t_N,t_{N-1})R_{a_{N-1}}U(t_{N-1},t_{N-2})\dots 
U(t_3,t_2)R_{a_2} U(t_2,t_1)R_{a_1} U(t_1,t_0)
$$
equal to the product of unitary evolution operators and positive operators presenting results of instantaneous fuzzy measurements, $\alpha=\{ a_1, a_2, \dots , a_{N-1} \}$. 

A more sophisticated way to describe CFM selectively is the nonlinear {stochastic Schr\"odinger equation} \cite{Belavkin89stochast,Diosi89stochast,Gisin89stochast}
\begin{eqnarray}
d|\psi\rangle &=& \left[
          -\frac{i}{\hbar} H - \kappa\big( A-\langle A\rangle\big)^2
          \right] |\psi\rangle dt
          + \sqrt{2\kappa} \big( A-\langle A\rangle\big)  |\psi\rangle dw, \nonumber\\ 
dw^2 &=& dt,
\label{stochEq}\end{eqnarray}
in which the influence of the environment is presented by the white noise $w$ according to the scheme

\quad

\framebox{{\bf System}} $\leftarrow$ {\bf Noise}\\\

\quad

Different stochastic equations were proposed which impose the same Lindblad equation (\ref{LindbladEq}) for the density matrix. It is evident that an additional criterion is necessary to choose correct selective description even if the nonselective description is known. The following approach is unambiguous. 

One may describe CFM selectively by the Schr\"odinger equation with a {\em complex Hamiltonian} \cite{GolubMen89qnd,MenOnoPre91,MenOnoPre93,Men-bk93} 
\begin{equation}
\frac{\partial}{\partial t} |\psi_t\rangle
  = \left[-\frac{i}{\hbar} H
  -\kappa \,\Big(A - a(t)\Big) ^2\right]\, |\psi_t\rangle 
\label{complHamEq}\end{equation}
where $a(t)$ is a readout of the continuous measurement of the observable $A$. The norm of the wave function at the end of the process determines the probability density of the given measurement readout $a(t)$. In this equation the influence of the environment is presented in a more concrete way, it is expressed through the information about the system which is recorded in the environment, according to the scheme

\quad

\begin{tabular}{llc} \label{info-scheme}
 				& $\rightarrow$ & {\bf Information}\\
\framebox{{\bf System}} & 			& $\downarrow$\\
				&	$\leftarrow$ & {\bf Influence}
\end{tabular}

\quad 

The complex-Hamiltonian equation (\ref{complHamEq}) is equivalent \cite{PreOnoTam96,AlbeverioKolSmol97} to  the stochastic equation (\ref{stochEq}) while the Lindblad equation (\ref{LindbladEq})  follows from any of them: 
\begin{tabbing}

\parbox{6.5cm}{\underline{Selective description}}
\hspace{0.4cm}\= \hspace{0.8cm}\=
\parbox{6.1cm}{\underline{Non-selective description}}\kill
\framebox{%
\parbox{6cm}{{\bf Complex-Hamiltonian equation}}
}
\> $\searrow$ \>
\\ 
\hspace{2cm}  $\updownarrow$
\> \>
\framebox{\parbox{3.5cm}{{\bf Lindblad equation}}}
\\ 
\framebox{%
\parbox{6cm}{{\bf Stochastic equation}}
}
\> $\nearrow$ \>
\end{tabbing}
However the complex-Hamiltonian equation may be derived from general principles with the help of restricted path integrals (see Sect.~\ref{SectRPI}), and this may be used as a criterion for the choice of the type of the stochastic equation. The relationship between different phenomenological approaches to theory of continuously measured systems may be presented by the following scheme:\\[1mm]

\begin{tabbing}

\parbox{6cm}{\underline{Selective description}}
\hspace{0.5cm}\= \hspace{0.8cm}\=
\parbox{4.5cm}{\underline{Nonselective description}}
\\
\\
\framebox{%
\parbox{6cm}{{\bf Path integral quantum mechanics}:\\Feynman 1948}
}
\> $\searrow$ \>
\\
\hspace{2cm}  $\downarrow$
\> \>
\hspace{2cm}
\\
\framebox{%
\parbox{6cm}{{\bf Restricted path integrals}:\\
             Mensky 1979}
}
\> $\rightarrow$ \>
\framebox{%
\parbox{4cm}{{\bf Influence functional}:\\ Feynman \& Vernon 1963}
}
\\
\hspace{2cm}  $\downarrow$
\> \>
\hspace{2cm}
\\
\framebox{%
\parbox{6cm}{{\bf Complex Hamiltonians}:\\
Golubtsova \& Mensky 1989\\
Mensky, Onofrio \& Presilla 1991}
}
\> $\searrow$ \>
\hspace{1.8cm}  $\downarrow$
\\ 
\hspace{2cm}  $\downarrow$
\> \>
\hspace{2cm}
\\ 
\framebox{%
\parbox{6cm}{{\bf Stochastic equation}:\\
Diosi 1989; Gisin 1989; Belavkin 1989}
}
\> $\rightarrow$ \>
\framebox{%
\parbox{4cm}{{\bf Master equation}:\\ Lindblad 1976}
}
\end{tabbing}

\section{Restricted path integrals (RPI)}
\label{SectRPI}

The complex-Hamiltonian Schr\"odinger equation (\ref{complHamEq}) may be derived in the framework of quantum theory of continuous measurements based on restricted path integrals (RPI) \cite{Feynman48,Men79a,Men79b,Men-bk93}. An advantage is that Eq.~(\ref{complHamEq}) is derived from general principles in a model-independent way. (However it may also be obtained from the consideration of a concrete measuring system, see below in Sect.~\ref{SectEnergy}). 

The evolution of a closed system is described by the evolution operator~$U_t $, 
$$
|\psi_t\rangle = U_t |\psi_0\rangle, \quad
\rho' = U_t \rho_0 U_t^{\dagger}
$$
presented in the Feynman version of quantum mechanics by a path integral: 
$$
U_T(q'',q')=\int d[q]\,e^{\frac{i}{\hbar}S[q]}
=\int d[p]d[q]\,
e^{\frac{i}{\hbar}\int_0^T (p\dot q - H(p,q,t))}
$$
\begin{figure}[ht]
\let\picnaturalsize=N
\def\picsize{5cm}
\def\picfilename{feynman.eps}
\ifx\nopictures Y\else{\ifx\epsfloaded Y\else\input epsf \fi
\let\epsfloaded=Y
\centerline{\ifx\picnaturalsize N\epsfxsize \picsize\fi \epsfbox{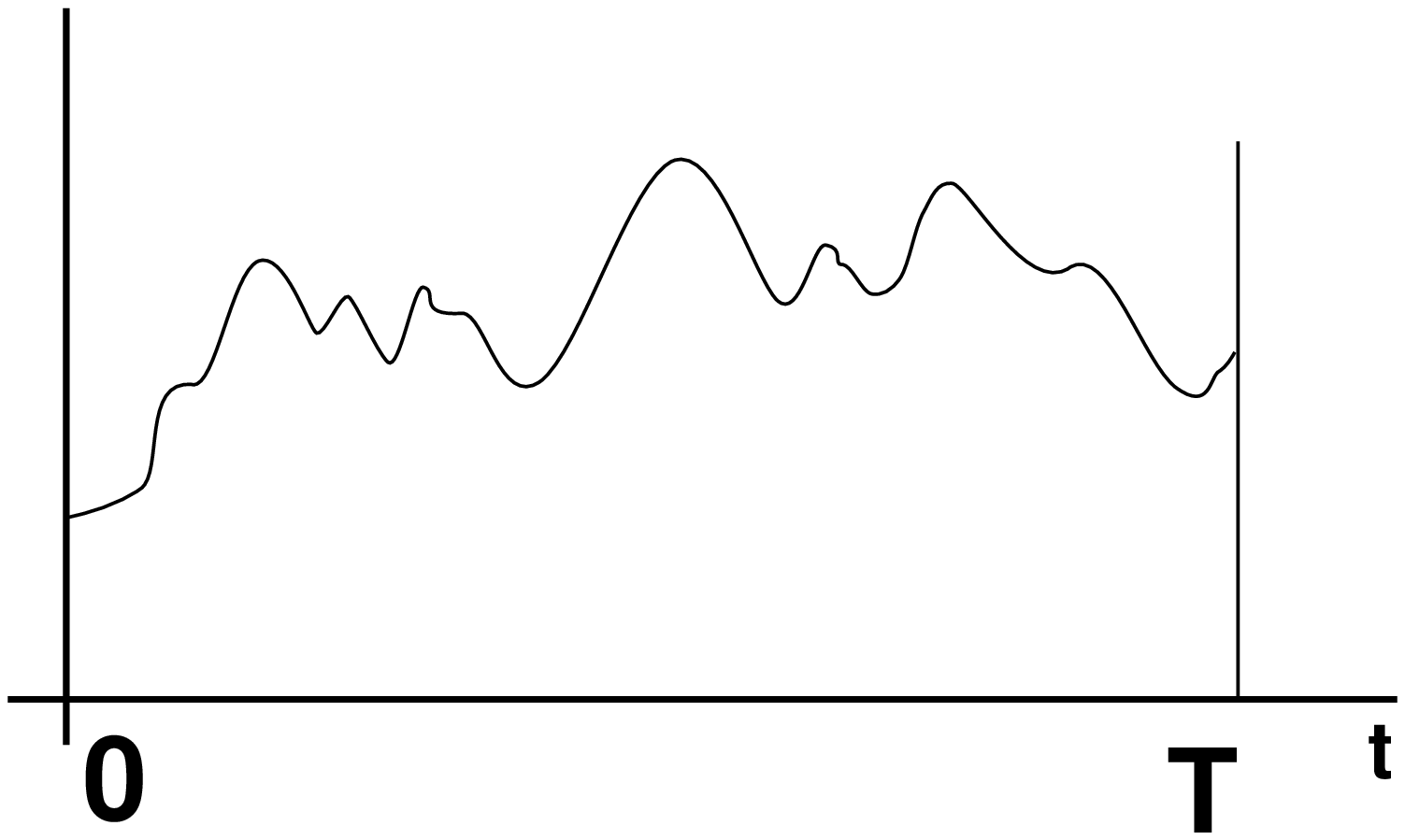}}}\fi
\label{fig-feynman}
\end{figure}

An ideology behind the path integral is that the total amplitude is a sum of the amplitudes $\exp (iS[q]/\hbar)$ corresponding to individual paths. In the case of a continuous measurement this must be the sum over those paths which are compatible with the information supplied by the measurement. This gives an integral restricted to a subset of paths, or the usual Feynman path integral but with a weight functional depending on the measurement readout $\alpha$: 
\begin{equation}
U_T^{\alpha}(q'',q')
=\int d[p]d[q]\,w_\alpha[p,q]\,
e^{\frac{i}{\hbar}\int_0^T (p\dot q - H(p,q,t))}
\label{RPI}\end{equation}
\begin{figure}[ht]
\let\picnaturalsize=N
\def\picsize{5cm}
\def\picfilename{rpi.eps}
\ifx\nopictures Y\else{\ifx\epsfloaded Y\else\input epsf \fi
\let\epsfloaded=Y
\centerline{\ifx\picnaturalsize N\epsfxsize \picsize\fi \epsfbox{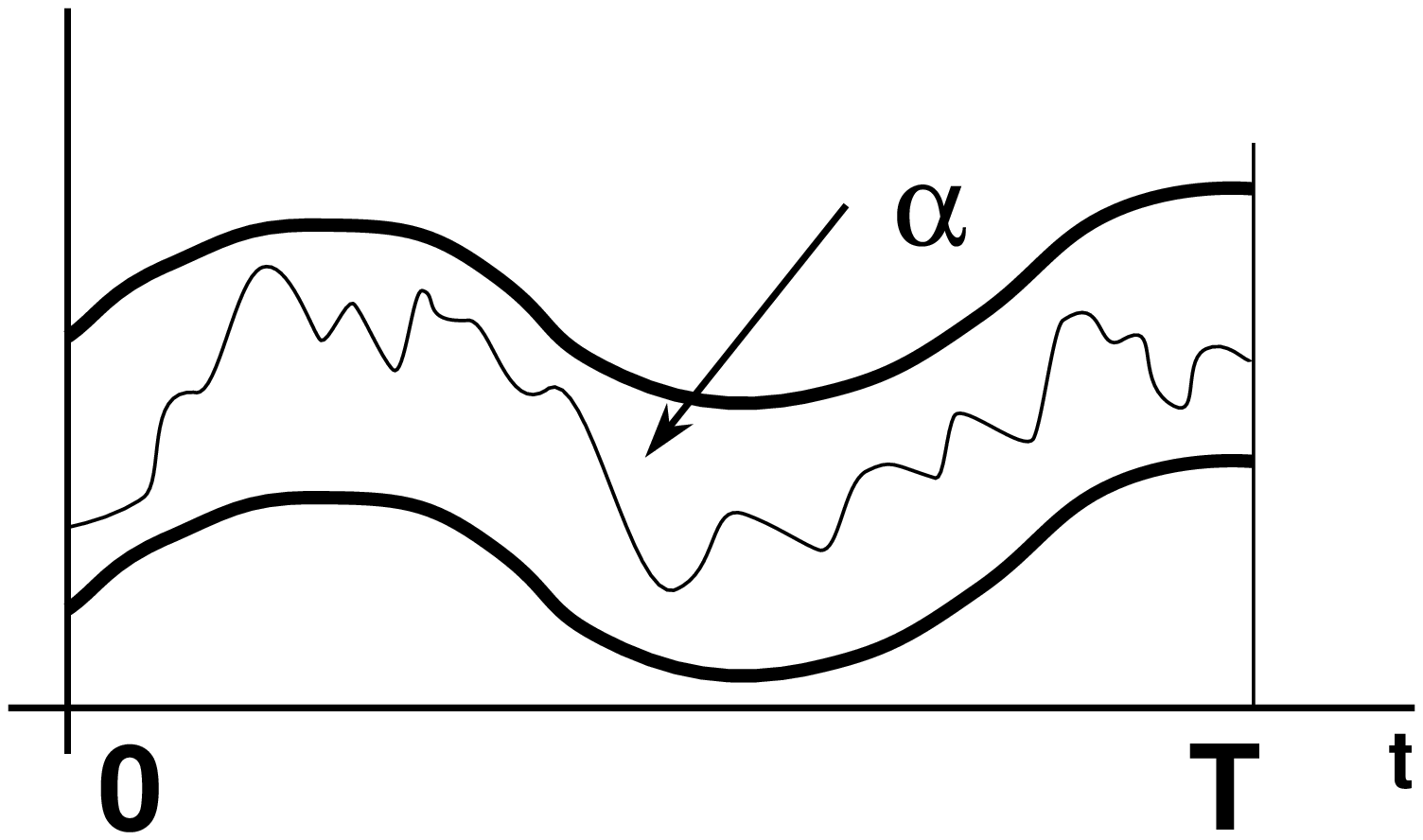}}}\fi
\label{fig-rpi}
\end{figure}

The evolution of the system (depending on the readout $\alpha$) is then 
$$
|\psi_T^{\alpha}\rangle = U_T^{\alpha} |\psi_0\rangle, \quad
\rho_T^{\alpha} = U_T^{\alpha} \rho_0 \left(
U_T^{\alpha}\right)^{\dagger}
$$
and the probability density of the readout $\alpha$ is $P(\alpha)=\mbox {Tr}\rho_T^{\alpha}=||\psi_T^{\alpha}||^2$ (where $T$ is the moment when the continuous measurement is over). A complete description of the evolution is given in this case by a family of {\em partial evolution operators} $U_t^{\alpha}$ corresponding to all possible measurement readouts $\alpha$. The set of paths corresponding to the given $\alpha$ may be called a {\em quantum corridor} in analogy with quantum trajectories of Carmichael \cite{Carmichael-bk93}. In the general case quantum corridors are described by weight functionals $w_\alpha$, i.e. they have ``unsharp boundaries". 

If we do not know the measurement readout or are not interested in it, we have to sum up the density matrix $\rho_T^{\alpha}$ over all readouts to obtain
$$
\rho_T = \int d\alpha\,  \rho_T^{\alpha}
=\int d\alpha\, U_T^{\alpha} \rho_0 \left( U_T^{\alpha}\right)^{\dagger}.
$$
The resulting density matrix $\rho_T$ is normalized if the following {\em generalized unitarity condition} is fulfilled: 
$$
\int d\alpha\,  \left( U_T^{\alpha}\right)^{\dagger}\, U_T^{\alpha} =
\mathbf 1.
$$

Consider a special type of continuous measurements, namely monitoring an observable. Let an observable $A=A(p,q,t)$ be monitored giving the readout $[a] = \{a(t)|0\le t \le T\}$. Then the corresponding set of paths is a corridor around $[a]$ of the width depending on the resolution of the measurement. 
\begin{figure}[ht]
\let\picnaturalsize=N
\def\picsize{5cm}
\def\picfilename{rpi-mon.eps}
\ifx\nopictures Y\else{\ifx\epsfloaded Y\else\input epsf \fi
\let\epsfloaded=Y
\centerline{\ifx\picnaturalsize N\epsfxsize \picsize\fi \epsfbox{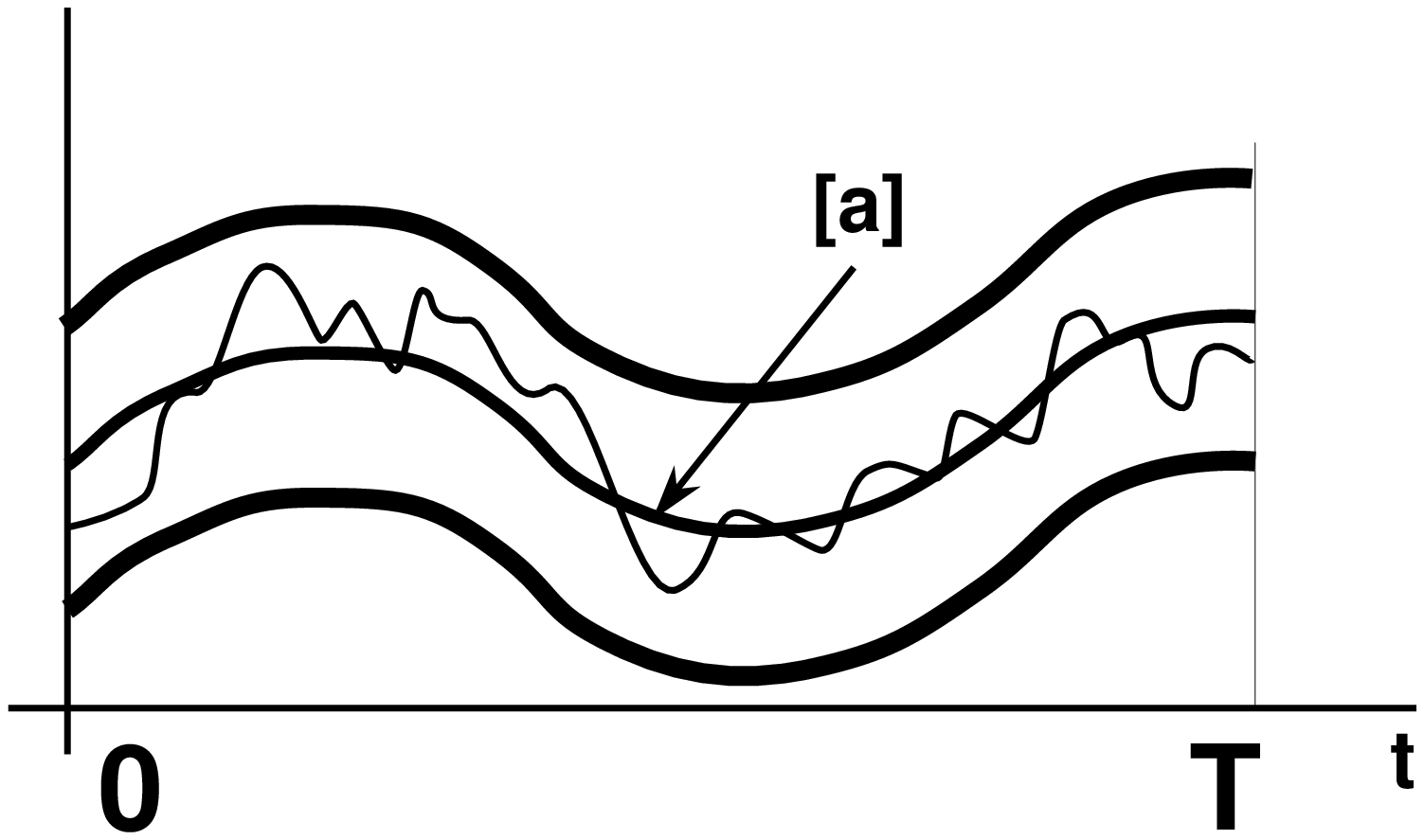}}}\fi
\label{fig-rpi-mon}
\end{figure}

The weight functional may be of the Gaussian form:
\begin{equation}
w_{[a]}[p,q] = \exp\left(
{-\kappa \int_0^T [ A(t) - a(t) ]^2\,dt }
\right)
\label{GaussWeight}\end{equation}
leading to the RPI of the form
\begin{eqnarray}
U_T^{[a]}(q'',q')=\\
\int d[p]\,d[q]\,\exp\left\{ \frac{i}{\hbar} \int_0^T \big(p\dot q
- H(p,q,t)\big)\,dt \right.\nonumber
- \left.\kappa \int_0^T \big(A(p,q,t)-a(t)\big)^2 dt
\right\}
\label{Umonitor}\end{eqnarray}
It is seen that this restricted path integral may be considered as an usual (nonrestricted) Feynman integral but with the {\em effective complex Hamiltonian} \cite{GolubMen89qnd} depending on the measurement readout $[a]$:
\begin{equation}
H_{[a]}\,(p,q,t) = H(p,q,t) - i\kappa\hbar \,\big( A(p,q,t) - a(t) \big)^2.
\label{effectHam}\end{equation}
The behavior of the continuously measured system is described then by the {\em effective Schr\"odinger equation} (\ref{complHamEq}). The solution of this equation is not normalized. Its norm at the final moment gives the probability density for the function $[a]$ to arise as a measurement readout. 

After summation over all possible measurement readouts (in the present case integration over all $[a]$) we have the density matrix $\rho(t)$ describing the measured system in a nonselective way and satisfying \cite{Men94MasterEq} the Lindblad equation (\ref{LindbladEq}). 

\begin{remark} {\rm It has been implicitly assumed in the preceding argument that monitoring is performed with the absolute resolution of time so that the number $a(t)$ is an estimate of the observable $A$ in the precisely known time moment $t$. In some cases the finite resolution of time (inertial properties of the measuring medium) must be taken into account. This may be done in the framework of the RPI approach \cite{Men97timeResol} and is equivalent to non-Markovian approximation in description of the process of measurement (decoherence).}
\end{remark}

\section{Measurement of energy}
\label{SectEnergy}

In 
\cite{OnoPreTam93En,OnoPreTam95En,AudMen97En,AuMenNam97scat,AuMenEnGen} 
the complex-Hamiltonian Schr\"odinger equation (\ref{complHamEq}) is applied to the case of continuous measurements of energy. Let the system Hamiltonian be $H_0 + V$ where $H_0$ presents a multilevel system, $V$ is a driving field inducing transitions between levels and the measured observable is $H_0$. Then the effective Schr\"odinger equation is 
\begin{equation}
\frac{\partial}{\partial t} |\psi_t\rangle
  = \left[-\frac{i}{\hbar} (H_0 + V)
  -\kappa \,\Big( H_0 - E(t)\Big) ^2\right]\, |\psi_t\rangle
\label{complHamEnEq}\end{equation}
where $\kappa$ characterizes resolution of the measurement and $[E]$ is its readout. 

This equation was explored in different situations, in most detail for a two-level system under resonance driving field. In \cite{OnoPreTam93En,OnoPreTam95En} it was shown that for an accurate enough measurement (big $\kappa$) quantum Zeno effect arises: the system is frozen i.e. Rabi oscillations are prevented. The apriori assumption $E(t)={\rm const}$ introduced in \cite{OnoPreTam93En,OnoPreTam95En} did not allow the authors to correctly consider inaccurate measurements. This was made in 
\cite{AudMen97En,AuMenNam97scat,AuMenEnGen}. 

In \cite{AudMen97En} it was shown that the continuous measurement of energy i)~provides a model for the process of decoherence, i.e. gradual approaching to the von Neumann projection on one of the energy levels and ii)~allows visualization (with restricted precision) of Rabi oscillations. In \cite{AuMenNam97scat} it was shown that a level transition may be monitored by the continuous measurement of energy without too strong back influence to its dynamics, and the concrete experimental setup was proposed for realization of such a measurement. In \cite{AuMenEnGen} a much larger class of realizations (by a long series of short weak interactions with a subsidiary system) was considered. The quadratic form of the imaginary term in the effective Hamiltonian in Eq.~(\ref{complHamEnEq}) turned out to be universal, independent of concrete features of interactions in the series (a quantum analogue for the limiting theorem of probability theory).

\section{Features of the RPI approach}

The RPI approach to continuous quantum measurements follows from general principles (Feynman form of quantum mechanics) and may also be derived from models of the measurement. This demonstrates its fundamental character. With this approach, Feynman quantum mechanics becomes close (self-sufficient) since it includes theory of measurements as well. The RPI approach reveals the role of information in the back influence of the measuring medium, or environment \cite{Men96InfoPrinc}. In fact, back influence described by the effective Schr\"odinger equation (\ref{complHamEq}) or by the partial evolution operator (\ref{RPI}) or (\ref{Umonitor}) depends only on the measurement readout i.e. on the information about the system recorded in its environment (see the upper scheme on page~\pageref{info-scheme}).

Let us note at the end that the RPI approach has similarity with the consistent-histories (CH) approach in quantum mechanics
 \cite{Griffiths84,Omnes90,GellMannHartle90}.
Quantum corridors in the RPI approach are similar to histories in CH approach. The difference is that the former approach deals with open (continuously measured) systems while the latter is developed for closed systems. In the measurement situation the closed system in the CH approach includes both the measured and measuring subsystems, while in the RPI approach only the measured system is included, while the influence of the environment is taken into account implicitly. This is why in the CH approach together with the given family of histories more coarse-grained families have to be considered and consistency  between them must be required. On the contrary, in the RPI approach the family of quantum corridors is fixed because it implicitly describes the given environment. Hence, no consistency condition has to be imposed.

\section{Conclusion}

\begin{itemize}
\item Continuous fuzzy measurements give models for gradual decoherence
\item The dynamics of a continuously measured (decohering) system depends only on the information recorded in its environment.
\item A quantum continuous fuzzy measurement allows one to monitor a quantum transition with moderate back influence on its dynamics.
\item Such a measurement may be realized by a long series of short weak interactions with a subsidiary system. The resulting evolution is correctly described by a quadratic imaginary term in the Hamiltonian.
\end{itemize}

\centerline{\bf ACKNOWLEDGEMENT}

The author is obliged to J.~Audretsch, V.~Namiot and H.-D.~Zeh for fruitful discussions. The work was supported in part by the Deutsche Forschungsgemeinschaft and Russian Foundation of Basic Research, grant 98-01-00161.


\end{document}